# Measurement Complexity of Quantum Compressed Sensing

Jianyong Hu and Wei Li

***Abstract*—Conventional compressed sensing (CS) has a measurement lower bound of $M = \Omega(K \log(N/K))$ under non-adaptive measurements. Recent experiments on quantum compressed sensing (QCS) have reported numbers of measurements below this classical lower bound. In this work, we establish lower bounds on the measurement complexity of QCS from an information-theoretic and quantum-physical perspective. QCS exploits quantum parallelism, enabling a unitary domain-alignment evolution to act on a superposition of all $N$ basis states and map the sparse basis onto the measurement basis in a one-to-one manner. Consequently, the locations of nonzero components are explicitly carried by measurement-index labels. If only basic statistical information about the $K$ nonzero components is required, the total number of effective index samples satisfies the coarse necessary condition $M_s = \Omega(K)$. Under independent single-index projective measurements, when the support is unknown, the nonzero probabilities satisfy $p_n = \Theta(1/K)$, and exact support recovery is required with a fixed success probability, the sampling-coverage requirement leads to $M_s = \Theta(K \ln K)$. If each nonzero amplitude coefficient is further required to achieve a coordinate-wise relative root-mean-square error $\varepsilon$, the optimal effective index sample complexity is $M_s = \Theta(K \ln K + K/\varepsilon^2)$, which reduces to $\Theta(K \ln K)$ for fixed $\varepsilon$. Multimode photon-number-resolving detection can reduce the number of experimental rounds but not the total effective index sample complexity. These results show that quantum parallelism is not only a core resource for quantum computing but can also play a transformative role in measurement science.**

## I. Introduction

A central question in sampling theory is how many measurements are fundamentally required to reconstruct a signal. For a signal of dimension $N$, the Nyquist–Shannon sampling theorem gives a linear scaling $M = \Omega(N)$, requiring the sampling rate to be at least twice the signal bandwidth. Under a finite-dimensional discretization over a fixed observation interval, this typically corresponds to a number of samples that grows linearly with the signal dimension [1], [2], [3]. This requirement is determined by the global bandwidth of the signal rather than by its intrinsic information structure. Consequently, even when the actual number of degrees of freedom is far smaller than $N$, uniform sampling must still be performed according to the full signal bandwidth. Compressed sensing (CS) breaks this linear constraint by exploiting signal sparsity in a known transform domain [4], [5], [6], [7], [8], [9], [10], [11]. When a signal is $K$-sparse in a sparse basis $\Psi$, the number of measurements under the non-adaptive linear measurement framework is lower bounded by $M = \Omega(K \log(N/K))$ [10], [11], [12], [13], [14], [15], [16], [17]. From an information-theoretic perspective, the logarithmic factor represents an unavoidable cost associated with the combinatorial entropy of the unknown support set [12], [13], [14], [15], [16], [17].

A natural question is whether this logarithmic scaling can be further reduced. Quantum compressed sensing (QCS) adopts a different measurement mechanism. The QCS protocol proposed in [18] exploits the parallelism of quantum superposition to prepare a probe state spanning all $N$ basis states simultaneously, and employs unitary domain-alignment evolution to map the sparse basis onto the measurement basis in a one-to-one manner. This measurement architecture creates a fundamental distinction between QCS and classical compressed sensing, and related concepts have also been applied to ultrawideband radio-frequency measurement, dynamic three-dimensional imaging, and long-range single-photon imaging [19], [20], [21]. In classical measurements, information about the locations of nonzero components is typically mixed into continuous-valued measurement outcomes and must be inferred from their numerical patterns. In QCS, by contrast, such location information is directly represented by the index labels of the measurement outcomes. However, the primary contribution of [18] was experimental validation. The theoretical origin of its quantum advantage, the mechanism by which QCS circumvents the classical information-theoretic lower bound, and the conditions under which this advantage holds have not yet been established within a rigorous and general information-theoretic framework.

This work aims to establish a rigorous information-theoretic foundation for the conditional quantum advantage of QCS. To distinguish different measurement resources, we use $M_s$ to denote the total number of effective index samples and $M$ to denote the number of experimental rounds. Under standard single-index projective measurement, each experimental repetition produces only one index outcome, and therefore $M = M_s$. With multimode photon-number-resolving detection, a single experimental round can produce multiple effective index samples, in which case the number of experimental rounds $M$ can be smaller than the total number of effective index samples $M_s$. We first consider the total number of effective index samples because it reflects the statistical resources required for the recovery task itself. Under this resource definition, the measurement complexity of QCS must be analyzed at different levels according to the recovery objective.

First, when only the total sample requirement for obtaining basic statistical information about the $K$ nonzero components is considered, one obtains $M_s = \Omega(K)$. This result indicates that the total number of effective index samples must grow at least linearly with the sparsity $K$, with no explicit dependence on the signal dimension $N$. However, this result is only a coarse general necessary condition and cannot be directly interpreted as a tight lower bound for exact recovery of an unknown support.

Second, under standard independent single-index projective measurements, if the support is unknown, the nonzero probabilities satisfy $p_n = \Theta(1/K)$, and exact support recovery is required with a fixed success probability, the recovery process is constrained by the sampling-coverage requirement. An unobserved index cannot be directly distinguished from a zero component. Moreover, under the unknown-support model with $N \geq 2K$, guessing among unobserved locations cannot eliminate the asymptotic cost associated with covering all nonzero indices. Therefore, the total number of effective index samples required for exact support recovery is $M_s = \Theta(K \ln K)$.

Third, if, in addition to exact support recovery, each nonzero probability or its corresponding amplitude is required to achieve a coordinate-wise relative root-mean-square error $\varepsilon$, probability estimation requires an additional $\Omega(K/\varepsilon^2)$ statistical samples. Since support identification and probability estimation can use the same set of index samples, the overall sample requirement is $M_s = \Theta(K \ln K + K/\varepsilon^2)$. For fixed $\varepsilon$, this reduces to $M_s = \Theta(K \ln K)$. When $\varepsilon$ is sufficiently small, the $K/\varepsilon^2$ term exceeds the $K \ln K$ term and becomes the dominant measurement cost.

The remainder of this paper discusses, in sequence, the conditions under which classical non-adaptive measurement bounds apply, the encoding and measurement protocol of QCS, the statistical lower bounds associated with support coverage and probability estimation, and the effect of multimode photon-number-resolving detection on the number of experimental rounds. We focus on measurement complexity under a specific signal-access model and index-readout model. The advantage of QCS relies on a known sparse basis, implementable domain alignment, and resolvable index-based readout. Accordingly, the results of this work should be understood as a conditional measurement-complexity advantage under the specified observation model, rather than as an unconditional quantum advantage over all classical measurement schemes.

## II. Information-Theoretic Lower Bound on the Number of Measurements for Classical Non-Adaptive CS

This section establishes the measurement-complexity benchmark for classical compressed sensing. We first specify the task constraints shared by classical CS and QCS, including a known sparse representation, an unknown support set, non-adaptive measurements, finite measurement resources, no additional prior information, and a common recovery criterion. We then take classical non-adaptive linear measurements as the benchmark for classical CS and derive a lower bound on the number of measurements using Fano's inequality under a finite per-measurement information budget or dimension-independent stable-embedding conditions [22], [23], [24].

### *A. Constraints*

The measurement schemes considered in this work satisfy the following five common constraints.

Constraint 1: Known sparse basis and unknown support set. The sparse basis Ψ is known before measurement, and the measurement scheme may exploit the structure of Ψ in its design. However, it cannot use the specific signal x, the sparse coefficient vector S, or the support set Ω. Apart from Ψ and K, no additional structural prior on the support, such as block structure, tree structure, or any other exploitable prior structure, is assumed.

Constraint 2: Non-adaptive measurements. The measurement scheme is completely determined before any data are acquired and does not depend on the specific unknown signal, the unknown support set, or any previously obtained measurement outcomes.

Constraint 3: Finite and well-defined measurement resources. The measurement outputs have finite quantization precision, signal-to-noise ratio, dynamic range, or number of resolvable outcomes. Let Ri denote the output of the i-th measurement, and let Ci denote the effective information budget provided by that measurement. The total information budget satisfies $I(\Omega; RM|D) \leq \Sigma$ Ci, where D denotes the known measurement design and noise conditions [22].

Constraint 4: No additional prior information. The support set Ω may be any K-element subset, and the nonzero coefficients may vary arbitrarily within the prescribed range. The measurement scheme does not exploit any signal-instance-specific prior information about the amplitudes, signs, phases, or locations of the nonzero components.

Constraint 5: A common recovery criterion. The core comparison between classical CS and QCS adopts the same support-recovery task: with the support set unknown, the support-recovery error probability is required to be no greater than δ. The measurement lower bound for classical CS and the support-recovery complexity of QCS are compared under this same criterion. Estimation of continuous-valued coefficients or probabilities constitutes a stronger, extended task and requires separate specifications of the error metric, target precision, noise, and quantization conditions. The complexities of the two schemes can be directly compared only when they are evaluated under the same precision requirements and resource model.

### *B. Measurement Lower Bound for Classical Non-Adaptive Linear CS*

For unstructured unknown sparse signals, linear measurements provide a natural universal benchmark under non-adaptive measurements, finite measurement resources, and a uniform worst-case recovery criterion. Under the same signal-access model, noise level, dynamic range, and total information-budget constraints, merely changing the nonlinear form of the observation does not by itself eliminate the combinatorial information requirement associated with the support set [12], [13], [14], [15], [16], [17], [22], [23], [24].

To establish a classical benchmark, we consider classical non-adaptive linear measurements. Let the signal $x \in \mathbb{R}^N$ satisfy

$$x = \Psi S = \sum_{n=1}^{N} s_n \psi_n , \tag{1}$$

where $\Psi \in \mathbb{R}^{N\times N}$ is a known orthogonal sparse basis and $S \in \mathbb{R}^N$ is a $K$-sparse coefficient vector. Its support set is denoted by $\Omega = \{n: s_n \neq 0\}$, with $|\Omega| = K$ and $K \ll N$. The classical non-adaptive linear measurement model is

$$Y = \Phi x + Z = \Phi\Psi S + Z , \tag{2}$$

where $\Phi \in \mathbb{R}^{M\times N}$ is the measurement matrix, which is fixed before the measurements are performed and is independent of the specific signal and the unknown support set. $Y$, $Z \in \mathbb{R}^M$ denote the measurement vector and the measurement-noise vector, respectively. After quantization or readout, let $R_i$ denote the actual output of the *i-th* measurement, and let $R^M = (R_1, \ldots, R_M)$ denote the complete set of outputs. We use $D$ to denote the known measurement design, including Ψ, Φ, the quantization rule, and the noise parameters. If Φ is randomly generated, we assume that $D$ is independent of the unknown support set Ω and is available to the decoder during recovery.

The uncertainty of the support set Ω is quantified by its combinatorial entropy [10], [22]. We regard Ω as a random variable uniformly distributed over all $K$-element subsets of the $N$ coordinates. The number of possible support sets is $B = C_N^K$, and hence its Shannon entropy is $H(\Omega) = \log_2 B$. For $1 \leq K \leq N/2$,

$$\log_2 B = \Theta\left(K \log_2 \frac{N}{K}\right). \tag{3}$$

To exclude the effect of amplitude estimation from the lower-bound derivation, we consider the following hard family of signals:

$$s_n(\Omega) = \begin{cases} \mu/\sqrt{K}, & n \in \Omega \\ 0, & n \notin \Omega \end{cases}, \tag{4}$$

where $\mu > 0$ is a constant independent of $N$ and $K$. This construction keeps the total signal energy independent of the support set and reduces the recovery task to support-set identification.

Suppose that the decoder produces an estimate $\hat{\Omega}$ of the support set from $(R^M, D)$. If its average error probability satisfies

$$P_e = \Pr\left(\Omega \neq \Omega\right) \leq \delta, \quad 0 \leq \delta < 1 , \tag{5}$$

then, by Fano's inequality [22], [23], [24],

$$H\left(\Omega \mid R^M, D\right) \leq h_2(\delta) + \delta \log_2(B-1), \tag{6}$$

where $h_2(\delta)$ is the binary entropy function. Since $D$ is independent of Ω,

$$H(\Omega \mid D) = H(\Omega) = \log_2(B). \tag{7}$$

Therefore,

$$\begin{aligned} I\left(\Omega;R^M\left|D\right.\right) &= H\left(\Omega\,|\,D\right) - H\left(\Omega\,|\,R^M,D\right) \\ &\geq \log_2 B - h_2(\delta) - \delta\log_2(B-1) \\ &\geq \left(1-\delta\right)\log_2 B - 1 \end{aligned} \tag{8}$$

On the other hand, let $C_i$ denote the conditional information budget provided by the $i$-th measurement. By the chain rule of mutual information [22],

$$I\left(\Omega;R^M\left|D\right.\right) = \sum_i^M I\left(\Omega;R^i\left|R^{i-1},D\right.\right) \leq \sum_i^M C_i\ . \tag{9}$$

Combining the above inequalities yields the following general information-budget necessary condition for classical non-adaptive linear measurements:

$$\sum_i^M C_i \geq \left(1-\delta\right)\log_2 B - 1\,. \tag{10}$$

Equation (10) is more general than a lower bound expressed directly in terms of the number of measurements. It shows that the total amount of information required for reliable recovery of an unknown support set must be at least of the same order as the combinatorial entropy of the support. If we further assume that the information budget of each classical measurement is bounded by a constant $C_0$ independent of $N$ and $K$, i.e., $C_i \leq C_0$, then (10) gives

$$MC_0 \geq \left(1-\delta\right)\log_2 B - 1\,. \tag{11}$$

For fixed $\delta < 1$ and $1 \leq K \leq N/2$, this yields [12], [13], [14], [15], [16], [17]

$$M = \Omega\left(K\log\frac{N}{K}\right). \tag{12}$$

Therefore, under classical non-adaptive linear measurements, an unknown support set, and a dimension-independent per-measurement information budget, even reliable identification of the locations of the $K$ nonzero components requires the number of measurements to satisfy $M = \Omega(K\log(N/K))$. This result is an algorithm-independent lower bound for support recovery and reflects the combinatorial information requirement associated with an unknown support set. If the amplitudes of the nonzero coefficients are further required to be estimated to a prescribed accuracy, the continuous amplitude space introduces additional information requirements. Consequently, the measurement lower bound for complete signal recovery cannot be lower than the result above and may increase further as the required estimation accuracy becomes more stringent; its specific form depends on the noise, quantization scheme, and error metric.

## III. Formalization of the QCS Protocol and Resource Model

This section provides a formal description of the standard QCS protocol and its resource model. We first introduce the four steps of quantum probe state preparation, signal encoding, domain-alignment evolution, and index-resolved projective measurement, and show how the sparse coefficients are mapped to measurement indices and their occurrence probabilities. We then verify that the protocol satisfies the common constraints specified in Section II and show that the fundamental distinction between QCS and classical CS lies in the measurement architecture and the information capacity of each measurement outcome, rather than in a relaxation of the recovery task or prior assumptions.

### *A. Four-Step QCS Protocol*

Following the signal model in Section II, let $x=\Psi S$, where $S$ is a $K$-sparse coefficient vector with an unknown support set $\Omega$, $|\Omega| = K$. The QCS protocol consists of the following four steps [18].

**Step 1: Quantum probe state preparation.** An initial quantum probe state $|\psi_p\rangle$ is prepared in an $N$-dimensional Hilbert space $\mathscr{H}_N$ as an $N$-dimensional superposition state,

$$\left|\psi_p\right\rangle = \frac{1}{\sqrt{N}}\sum_{n=1}^N \left|\psi_n\right\rangle. \tag{13}$$

The squared coefficients of the basis states are uniformly distributed,

$$\left|\left\langle\psi_n\left|\psi_p\right.\right\rangle\right|^2 = \frac{1}{N}, \qquad n = 1,2,\ldots,N\ . \tag{14}$$

This uniformity implies that the probe state has no preference for any sparse-basis vector and therefore contains no prior information about the signal or its support set.

**Step 2: Linear signal-to-quantum-state mapping.** A signal-controlled unitary evolution $\hat{U}_x$ is constructed to linearly encode the signal $x$ into the quantum probe state, yielding

$$\left|\psi_x\right\rangle = U_x\left|\psi_p\right\rangle = \sum_{n=1}^{N}\sqrt{\frac{x_n}{\|X\|_1}}\left|\psi_n\right\rangle, \tag{15}$$

where $\|X\|_1 = \sum_{n=1}^{N}|x_n|$ is the normalization factor.

**Step 3: Domain-alignment evolution.** A physically realizable unitary operator $\hat{U}_e$ is applied,

$$U_e = \sum_{n=1}^{N}\left|\omega_n\right\rangle\left\langle\psi_n\right|, \tag{16}$$

to align the sparse domain with the measurement domain, where $|\omega_n\rangle$ denotes a measurement-basis state. The evolved quantum state is

$$\left|\psi_{out}\right\rangle = \sum_{n=1}^{N}\sqrt{\frac{s_n}{\|S\|_1}}\left|\omega_n\right\rangle, \tag{17}$$

where $\|S\|_1 = \sum_{n=1}^{N}|s_n|$ is the normalization factor. This evolution establishes, within a single coherent operation, a direct correspondence between the sparse coefficients and the indices of the measurement basis, without increasing the number of online measurements. Its implementation, however, requires additional quantum-control resources, including a control structure associated with the known sparse basis and the ability to realize the corresponding unitary evolution; implementation details are given in [18]. In this work, we focus on the number of online measurements and do not consider the engineering complexity of the measurement apparatus or control architecture. If the signal $x$ is $K$-sparse, then after domain-alignment evolution, the superposition state has nonzero coefficients on only $K$ basis states, while the remaining $N$-$K$ basis-state coefficients are zero.

**Step 4: Projective measurement and statistical estimation.** Projective measurements are performed on $|\psi_{\text{out}}\rangle$ in the measurement basis $\{|\omega_n\rangle\}$. According to the Born rule [25], the probability of obtaining index $n$ is

$$p_n = \left|\left\langle\omega_n\middle|\psi_{out}\right\rangle\right|^2 = \eta\frac{|s_n|}{\|S\|_1}, \tag{18}$$

Here, $\eta$ denotes a global proportionality factor, and $s_n$ is the sparse coefficient. After $M$ independent projective measurements, each experimental repetition produces one index $y_m \in \{1, \ldots, N\}$, yielding the outcome sequence $Y = (y_1, \ldots, y_M)$. Under standard single-index projective measurement, the total number of index samples equals the number of experimental rounds $M$. The estimated probability is obtained by counting statistics as

$$p_n = \frac{k_n}{M}, \tag{19}$$

where $k_n$ is the number of occurrences of $y_m = n$. The magnitude of the sparse coefficient is then estimated as $\hat{s}_n = \hat{p}_n\|S\|_1$, and the original signal is reconstructed through the linear inverse transform $\hat{x} = \Psi\hat{S}$.

The QCS recovery task can be decomposed into two subtasks. Support identification determines which indices appear in the measurement sequence $Y$. Since $P(\omega_n) = 0$ for $n \notin \Omega$, the occurrence of any index $n$ at least once is sufficient to confirm that $n \in \Omega$. Amplitude estimation, once the support is known, uses the occurrence frequency $k_n/M$ of each index to estimate $|s_n|/\|S\|_1$. These two tasks are carried by different statistical features of the measurement data: support information is explicitly represented by the support pattern of the observed indices, whereas amplitude information is encoded in their occurrence frequencies.

### *B. QCS Satisfies the Common Constraints*

This subsection shows that QCS satisfies the five common constraints introduced in Section II-A. These constraints are imposed to ensure a common recovery task, prior assumptions, and resource accounting, rather than to require classical CS and QCS to adopt the same measurement form. Classical CS employs non-adaptive linear measurements, whereas QCS employs quantum-state encoding, unitary domain-alignment evolution, and index-resolved projective measurement.

**Constraint 1: Known sparse basis and unknown support set.** The domain-alignment evolution operator in QCS is constructed from the known sparse basis and the measurement basis. Therefore, constructing $\hat{U}_e$ requires prior knowledge of the sparse basis $\Psi$, but does not require knowledge of the specific signal $x$, the nonzero sparse coefficients $S$, or the support set $\Omega$. The same $\hat{U}_e$ applies to all signals that are sparse in $\Psi$. Knowledge of the sparse basis does not imply knowledge of the support set. The support set must still be identified from the index samples generated by projective measurements. Therefore, QCS satisfies the constraint of a known sparse basis and an unknown support set.

**Constraint 2: Non-adaptive measurements.** The state-preparation procedure, the domain-alignment operator $\hat{U}_e$, the measurement basis $\{|\omega_n\rangle\}$, and the projective-measurement rule in QCS are all determined before any measurement outcome is obtained. During the measurement process, these settings are not adjusted according to previous measurement outcomes, support estimates, or the specific signal instance. The unknown signal determines the output quantum state and its measurement probabilities, which is part of the data-generation process itself and does not constitute adaptive feedback. Non-adaptivity requires the measurement configuration to remain fixed; it does not require the measurement outcomes to be independent of the unknown signal. When repeated measurements are performed on the same unknown signal, each experimental round applies the same state preparation, domain alignment, and projective measurement operations. Therefore, QCS is a non-adaptive quantum measurement scheme. It satisfies the common non-adaptivity constraint, although its observation model differs from the classical linear measurement relation.

**Constraint 3: Finite and well-defined measurement resources.** Under standard single-index projective measurement, each successful measurement produces one index outcome. A single outcome belongs to a finite alphabet of size $N$, and fully specifying one index requires at most $\log_2 N$ bits [22], [26]. Therefore, the resource associated with each QCS measurement outcome is finite and explicitly quantifiable, but it does not satisfy a constant information-budget bound independent of $N$. This difference between QCS and classical finite-precision scalar measurements is reflected precisely in their different output spaces and per-measurement information capacities.

**Constraint 4: No additional prior information.** Apart from the known sparse basis $\Psi$ and sparsity level $K$, QCS does not exploit any additional information about the specific signal instance. The domain-alignment operator $\hat{U}_e$ is independent of the actual support set $\Omega$ and of the specific values of the nonzero sparse coefficients. Before measurement, neither the identities of the nonzero indices nor the probabilities with which individual indices will occur are known.

**Constraint 5: A common recovery criterion.** For the core comparison, both classical CS and QCS are required to exactly recover the locations of the nonzero components when the support set is unknown, with the support-recovery error probability no greater than $\delta$. QCS estimates the support set from the index samples generated by projective measurements and can therefore be evaluated using the same success-probability criterion as classical CS. Based on this criterion, Section IV derives the support-recovery complexity of QCS and further considers probability estimation under a coordinate-wise relative error requirement $\varepsilon$. The latter constitutes a stronger, extended task and is not used for direct comparison with the classical support-recovery lower bound.

In summary, QCS satisfies the five common constraints considered in this work: a known sparse basis, an unknown support set, a non-adaptive measurement configuration, finite and well-defined measurement resources, no additional signal-instance-specific prior information, and a common recovery criterion. The distinction between QCS and classical CS lies in their observation models and signal-access mechanisms, rather than in any relaxation of these common task conditions. Under this common framework, the following sections analyze the measurement complexity of classical non-adaptive linear measurements and QCS index measurements, respectively.

## IV. Information-Theoretic Lower Bounds for QCS Under Independent Single-Index Projective Measurements

This section characterizes the measurement complexity of QCS under the independent single-index projective-measurement model at three levels of recovery. We first establish the necessary condition $M_s = \Omega(K)$ for basic statistical estimation at fixed accuracy. We then show that, when the support is unknown, the nonzero probabilities satisfy $p_n = \Theta(1/K)$, and exact support recovery is required with a fixed success probability, the required number of effective index samples is $M_s = \Theta(K \ln K)$. This second-level result adopts the same support-recovery criterion as that used in Section II and therefore serves as the core benchmark for comparing the measurement complexity of QCS with that of classical CS. Finally, we introduce a coordinate-wise relative error requirement $\varepsilon$ and obtain the joint recovery complexity $M_s = \Theta(K \ln K + K/\varepsilon^2)$, while specifying the conditions under which each result applies.

### A. Level I: Basic Statistical Lower Bound for QCS

We first consider independent single-index projective measurements. In each experimental repetition, a quantum state is independently prepared and one measurement index is produced. Multi-index parallel readout and photon-number-resolving detection are not considered in this subsection. Let $M_s$ denote the total number of effective index samples and $M$ the number of experimental rounds. Under the present model, each experimental repetition produces one effective index sample, and hence $M = M_s$.

In the QCS model considered here, the quantum state after domain alignment is

$$\left|\psi_{out}\right\rangle = \sum_{n=1}^{N} \sqrt{p_n}\left|\omega_n\right\rangle , \tag{20}$$

where $p_n = s_n/\|S\|_1$, $\sum_{n=1}^{N} p_n = 1$. A projective measurement of $|\psi_{\text{out}}\rangle$ produces an index-valued random variable $J \in \{1, \ldots, N\}$ with probability distribution $\Pr(J = n) = p_n$. After $M_s$ independent state preparations and measurements, we obtain the index sequence $J_1, J_2, \ldots, J_{Ms}$. Define

$$k_n = \sum_{m=1}^{M_s} 1\left\{J_m = n\right\} . \tag{21}$$

The count vector then follows

$$\left(k_1,...,k_N\right) \sim Multinomial\left(M_s; p_1,..., p_N\right) , \tag{22}$$

and the empirical probability is

$$p_n = \frac{k_n}{M_s} , \tag{23}$$

Therefore, under standard single-index projective measurement, the statistical problem in QCS is equivalent to drawing $M_s$ independent samples from an unknown discrete probability distribution with support size $K$ [22].

It is important to note that although each measurement produces only one index, that index can take one of $N$ possible values. Fully specifying a single index generally requires $\log_2 N$ bits. Thus, the resource associated with a single measurement is finite, but its information capacity is not bounded by a constant independent of $N$. The $\Omega(K)$ lower bound derived below therefore arises from the statistical sampling requirement of probability estimation, rather than from an assumption of a fixed constant number of bits per measurement.

To obtain a uniform result that depends only on $K$, we consider the balanced probability class

$$\mathcal{P}_K\left(c,C\right) = \left\{ \mathrm{p} : \left|\Omega\right| = K, \frac{c}{K} \le p_n \le \frac{C}{K}, n \in \Omega \right\}, \tag{24}$$

where $0 < c < 1 < C$ are constants independent of $N$ and $K$. This condition requires all nonzero probabilities to be of order $1/K$. Without a lower bound on the nonzero probabilities, one nonzero component could become arbitrarily close to zero, in which case no finite sample complexity depending only on $K$ could be guaranteed. The balanced-probability condition is therefore a necessary condition for the lower bounds derived below.

We now establish the first-level lower bound. For the moment, assume that the support set is known and require only a fixed constant-level relative accuracy for the nonzero probabilities. Since the known-support problem is easier than exact recovery with an unknown support, any lower bound obtained for this simplified task also applies to the original problem.

Fix the support set as $\Omega = \{1, \ldots, K\}$, and let $0 < \alpha < 1/4$ be a constant independent of $K$. Construct two probability distributions $p^+$ and $p^-$ with

$$p_1^+ = \frac{1+\alpha}{K}, \qquad p_1^- = \frac{1-\alpha}{K} , \tag{25}$$

and

$$p_j^+ = \frac{1-p_1^+}{K-1}, \qquad p_j^- = \frac{1-p_1^-}{K-1}, \qquad j = 2,...,K. \tag{26}$$

For sufficiently small $\alpha$, both $p^+$ and $p^-$ belong to $\mathcal{P}_K(c, C)$. The difference between the two distributions at the first nonzero coordinate is $2\alpha/K$. Relative to the probability scale$1/K$, this difference is of constant order:

$$\frac{p_1^+ - p_1^-}{p_1^\pm} = \Theta\left(\alpha\right). \tag{27}$$

Therefore, if an estimator can estimate $p_1$ with a fixed relative accuracy, it can distinguish $p^+$ from $p^-$ with a fixed success probability. The Kullback–Leibler divergence between the two single-sample distributions is

$$D_{KL}\left(\mathrm{p}^+ \middle\| \mathrm{p}^-\right) \le \sum_{j=1}^{K} p_j^+ \ln \frac{p_j^+}{p_j^-} . \tag{28}$$

This quantity measures the difficulty of determining, from a single sample, whether the data are drawn from $p^+$ or $p^-$. Since the two distributions have the same conditional distribution outside the first coordinate, their single-sample Kullback–Leibler divergence is equal to that between the corresponding Bernoulli distributions [22], [24], [27]:

$$D_{KL}\left(\mathrm{p}^+ \middle\| \mathrm{p}^-\right) = D_{KL}\left(Bern\left(\frac{1+\alpha}{K}\right) \middle\| Bern\left(\frac{1-\alpha}{K}\right)\right). \tag{29}$$

Using the upper bound on the Bernoulli divergence,

$$D_{KL}\left(\mathrm{p}^+ \middle\| \mathrm{p}^-\right) \le \frac{\left(p_1^+ - p_1^-\right)}{p_1^-\left(1-p_1^-\right)}. \tag{30}$$

For $K \ge 2$ and $\alpha \le 1/4$,

$$p_1^- = \frac{1-\alpha}{K} \ge \frac{3}{4K}. \tag{31}$$

Moreover, $1 - p_1^- \ge 1/2$. Therefore,

$$D_{KL}\left(\mathrm{p}^+ \middle\| \mathrm{p}^-\right) \le \frac{16\alpha^2}{K}. \tag{32}$$

For $M_s$ independent samples, the Kullback–Leibler divergence is additive:

$$D_{KL}\left(\left(\mathrm{p}^+\right)^{\otimes M_s} \middle\| \left(\mathrm{p}^-\right)^{\otimes M_s}\right) = M_s D_{KL}\left(\mathrm{p}^+ \middle\| \mathrm{p}^-\right) \le \frac{16 M_s \alpha^2}{K}. \tag{33}$$

By Pinsker's inequality [24], [28],

$$\left\| \left(\mathrm{p}^+\right)^{\otimes M_s} - \left(\mathrm{p}^-\right)^{\otimes M_s} \right\|_{TV} \le \sqrt{\frac{1}{2} D_{KL}\left(\left(\mathrm{p}^+\right)^{\otimes M_s} \middle\| \left(\mathrm{p}^-\right)^{\otimes M_s}\right)}. \tag{34}$$

To distinguish the two distributions with a fixed success probability, the total variation distance must be of constant order. It is therefore necessary that

$$\frac{M_s \alpha^2}{K} = \Omega(1), \qquad M_s = \Omega\left(\frac{K}{\alpha^2}\right). \tag{35}$$

When $\alpha$ is fixed, this gives

$$M_s = \Omega(K). \tag{36}$$

This result shows that even when the support set is already known, obtaining a fixed relative accuracy for $K$ nonzero probabilities of order $1/K$ requires a number of effective index samples that grows at least linearly with $K$. Exact recovery with an unknown support cannot have a lower sample complexity.

It should be emphasized that the result $M_s = \Omega(K)$ obtained in this subsection is only a basic statistical lower bound at fixed accuracy and applies under the following conditions:

1) Only standard independent single-index projective measurements are considered, with each experimental repetition producing one index.
2) Square-root amplitude encoding is used, such that the measurement probabilities satisfy $p_n = s_n/\|S\|_1$.
3) The nonzero probabilities satisfy the balanced condition $p_n = \Theta(1/K)$.
4) Exact recovery of the unknown support is not required for this result.
5) The result applies only to normalized amplitude probabilities and does not include recovery of coefficient signs, phases, or the overall scale.

Therefore, the $\Omega(K)$ scaling cannot be directly interpreted as a tight lower bound for exact recovery of an unknown support. Exact support recovery additionally requires analysis of the index-coverage problem. When a coordinate-wise relative error $\varepsilon$ is further imposed, the additional sample cost associated with probability-estimation accuracy must also be analyzed. These two issues are addressed separately in the following subsections.

### B. Level II: Lower Bound for Exact Recovery of an Unknown Support

We continue to consider independent single-index projective measurements. Each experimental repetition produces an index $J_m \in \{1, \ldots, N\}$, with $\Pr(J_m = n) = p_n$. The support set of the unknown probability distribution $\mathbf{p} = (p_1, \ldots, p_N)$ is $\Omega = \{n: p_n>0\}$, $|\Omega| = K$. Assume that the nonzero probabilities belong to the balanced probability class defined in Section IV-A, namely,

$$\frac{c}{K} \le p_n \le \frac{C}{K}, \qquad n \in \Omega, \tag{37}$$

where $c$ and $C$ are positive constants independent of $N$ and $K$. The objective of this subsection is to exactly recover the unknown support set from the index sequence $J^{Ms} = (J_1, \ldots, J_{Ms})$. For a fixed failure probability $0 < \delta < 1/2$, the recovery algorithm is required to satisfy, uniformly over all support sets and probability distributions in the prescribed signal class,

$$\Pr_p\left(\Omega \ne \Omega\right) \le \delta. \tag{38}$$

1) *Uniform-Probability Subclass*

To establish a lower bound for uniform recovery, it is sufficient to select a difficult subclass from the balanced probability class. Consider the uniform nonzero-probability model

$$p_n = \begin{cases} \dfrac{1}{K}, & n \in \Omega, \\ 0, & n \notin \Omega. \end{cases} \tag{39}$$

Let the support set $\Omega$ be uniformly distributed over all $K$-element subsets, i.e.,

$$\Pr\left(\Omega = A\right) = \frac{1}{\binom{N}{K}}, \qquad |A| = K. \tag{40}$$

If an algorithm can correctly recover every support set with probability at least $1 - \delta$, then its average success probability under the above uniform prior must also be at least $1 - \delta$. Therefore, an average-error lower bound established for this subclass also implies a uniform worst-case lower bound for the original problem.

Let $\mathcal{O}_{Ms} = \{J_1, \ldots, J_{Ms}\}$ denote the set of distinct indices that have been observed, and let $D_{Ms} = |\mathcal{O}_{Ms}|$ denote the number of distinct observed indices. Clearly, $D_{Ms} \le K$. If $D_{Ms} = d$ distinct indices have been observed, all of them must belong to the true support set. However, when $d < K$, there remain $K - d$ nonzero locations that have not yet been observed. Conditioned on the complete measurement data, all $K$-element support sets containing $\mathcal{O}_{Ms}$ have the same posterior probability. The number of candidate support sets is therefore

$$\binom{N-d}{K-d}. \tag{41}$$

Hence, regardless of the recovery algorithm employed, its conditional success probability is bounded by

$$\Pr\left(\Omega = \Omega \middle| J^{M_s}\right) \le \frac{1}{\binom{N-D_{M_s}}{K-D_{M_s}}}. \tag{42}$$

When $D_{Ms} < K$,

$$\binom{N-D_{M_s}}{K-D_{M_s}} \ge N-K+1. \tag{43}$$

The overall success probability therefore satisfies

$$\Pr\left(\Omega = \Omega\right) \le \Pr\left(D_{M_s} = K\right) + \frac{\Pr\left(D_{M_s} < K\right)}{N-K+1}. \tag{44}$$

This inequality explicitly accounts for the possibility that the algorithm may guess the unobserved support locations. It does not simply assume that the support can be recovered only after all nonzero indices have been observed. Instead, it shows that when some nonzero indices remain unobserved, the algorithm must guess among a large number of equally likely candidate locations, resulting in a low probability of successful recovery.

2) *Number of Samples Required to Cover All Nonzero Indices*

Define the number of unobserved nonzero indices as $U_{Ms} = K - D_{Ms}$. For each $n \in \Omega$, define the missing-index indicator

$$I_n = \begin{cases} 1, & n \notin \mathcal{O}_{M_s} \\ 0, & Others \end{cases} . \tag{45}$$

Then

$$U_{M_s} = \sum_{n\in\Omega} I_n . \tag{46}$$

Under the uniform-probability model, the probability that any given index remains unobserved after $M_s$ samples is [29], [30], [31]

$$\Pr\left(I_n = 1\right) = \left(1 - \frac{1}{K}\right)^{M_s} . \tag{47}$$

Therefore,

$$\mathbb{E}\left[U_{M_s}\right] = K\left(1 - \frac{1}{K}\right)^{M_s} . \tag{48}$$

For two distinct indices $n \neq j$, the probability that both remain unobserved is

$$\mathbb{E}\left[I_n I_j\right] = \left(1 - \frac{2}{K}\right)^{M_s} . \tag{49}$$

Since

$$1 - \frac{2}{K} \le \left(1 - \frac{1}{K}\right)^2 , \tag{50}$$

we have

$$\mathbb{E}\left[I_n I_j\right] \le \mathbb{E}\left[I_n\right]\mathbb{E}\left[I_j\right] . \tag{51}$$

Thus, the covariances between distinct missing-index indicators are nonpositive, and consequently [31],

$$Var\left(U_{M_s}\right) \le \sum_{n\in\Omega} Var\left(I_n\right) \le \mathbb{E}\left[U_{M_s}\right] . \tag{52}$$

Now suppose that there exists a fixed constant $0 < \eta < 1$ such that $M_s \le (1 - \eta) K \ln K$. Using

$$\ln\left(1 - \frac{1}{K}\right) = -\frac{1}{K} + O\left(\frac{1}{K^2}\right) . \tag{53}$$

we obtain

$$\begin{aligned} \mathbb{E}\left[U_{M_s}\right] &= K\left(1 - \frac{1}{K}\right)^{M_s} \\ &= K \exp\left[M_s \ln\left(1 - \frac{1}{K}\right)\right] . \\ &\ge K^{\eta} + o(1) \end{aligned} \tag{54}$$

Hence, $\mathbb{E}[U_{Ms}] \to \infty$ as $K \to \infty$. By Chebyshev's inequality [31],

$$\begin{aligned}\Pr\left(D_{M_s}=K\right)&=\Pr\left(U_{M_s}=0\right)\\&\le\Pr\left(\left|U_{M_s}-\mathbb{E}\left[U_{M_s}\right]\right|\ge\mathbb{E}\left[U_{M_s}\right]\right)\\&\le\frac{Var\left(U_{M_s}\right)}{\mathbb{E}\left[U_{M_s}\right]^2}\\&\le\frac{1}{\mathbb{E}\left[U_{M_s}\right]}\to 0\end{aligned}\qquad .\tag{55}$$

If we further assume $2 \le K \le N/2$, then $N - K + 1 \ge K + 1$, and therefore

$$\frac{\Pr(D_{M_s}<K)}{N-K+1}\le\frac{1}{K+1}\to 0\,.\tag{56}$$

Combining these results with the preceding upper bound on the recovery success probability gives $\Pr\left(\hat{\Omega}=\Omega\right)\to 0$. That is, when $M_s \le (1 - \eta)\, K \ln K$, the average success probability of any recovery algorithm tends to zero and therefore cannot satisfy a fixed success-probability requirement uniformly over all support sets. Consequently, exact recovery of an unknown support requires $M_s = \Omega(K \ln K)$. More precisely, for any fixed $0 < \eta < 1$, when $K$ is sufficiently large, $M_s \le (1 - \eta)\, K \ln K$ is insufficient to achieve exact support recovery with a fixed success probability. This shows that the leading constant in the asymptotic lower bound is at least 1.

3) *Tightness of the Lower Bound*

The above lower bound can be achieved, up to the same asymptotic order, by a simple observed-support estimator. Define $\hat{\Omega}_{obs}=\left\{n:k_n>0\right\}$ .Under the ideal index-measurement model, the probabilities outside the support set are zero, so this estimator does not produce false indices. Recovery fails only if at least one nonzero index is not observed. Let $p_{\min} = \min(p_n)$, $n \in \Omega$. By the union bound [31],

$$\begin{aligned}\Pr\left(\Omega_{obs}\ne\Omega\right)&\le\sum_{n\in\Omega}(1-p_n)^{M_s}\\&\le\sum_{n\in\Omega}\exp(-M_s p_n)\\&\le K\exp(-M_s p_{min})\end{aligned}\quad .\tag{57}$$

For the balanced probability class, $p_{\min} \ge c/K$. Therefore,

$$\Pr\left(\Omega_{obs}\ne\Omega\right)\le K\exp\left(-\frac{cM_s}{K}\right).\tag{58}$$

To ensure that the failure probability does not exceed $\delta$, it is sufficient that $K \exp(-cM_s/K) \le \delta$, or equivalently,

$$M_s\ge\frac{K}{c}\ln\left(\frac{K}{\delta}\right).\tag{59}$$

For fixed $\delta$, this gives $M_s = O(K \ln K)$. Combining this upper bound with the preceding lower bound yields the exact support-recovery complexity under standard independent single-index projective measurements:

$$M_s=\Theta\left(K\ln K\right).\tag{60}$$

The ln $K$ factor arises from repeated observations in sampling with replacement. Although the $K$ nonzero locations correspond to only $K$ effective indices, some indices are sampled repeatedly. Consequently, covering all $K$ nonzero locations requires a number of samples on the order of the coupon-collector problem [29], [30]. The absence of an explicit dependence on $N$ arises because each measurement outcome directly provides a location label taking values in $\{1, \ldots, N\}$. The dimension $N$ determines the length of the index label and the required readout resolution; once these resources are available, however, the statistical bottleneck for exact support recovery is covering the $K$ effective indices rather than searching over all $N$ locations.

The results of this subsection hold under the following conditions: the unknown support set may be any $K$-element subset, with $2 \le K \le N/2$; the nonzero probabilities satisfy $p_n = \Theta(1/K)$; the measurement outcomes are independent single-index samples; and the support-recovery failure probability $\delta$ is a fixed constant. If $p_{\min} \ll 1/K$, rare nonzero components require more samples. If $\delta$ decreases with $K$, the confidence factor $\ln(K/\delta)$ must be retained. Structured supports, additional prior information, adaptive measurements, coherent queries, and noise or dark counts are outside the model considered in this subsection and may alter the measurement complexity.

### *C. Level III: Support Recovery with Coordinate-Wise Accuracy Requirements*

In addition to exact recovery of the unknown support set, we now further require coordinate-wise relative-accuracy estimation of each nonzero probability. We continue to consider independent single-index projective measurements and adopt the balanced probability class

$$\mathcal{P}_K(c,C)=\left\{\mathrm{p}:|\Omega|=K,\frac{c}{K}\le p_n\le\frac{C}{K},n\in\Omega\right\},\tag{61}$$

where $\Omega = \{n: p_n > 0\}$, and $c$ and $C$ are positive constants independent of $N$ and $K$. Suppose that $M_s$ measurements produce independent index samples $J_1, \ldots, J_{Ms}$, with $\Pr(J_m = n) = p_n$. The recovery algorithm outputs a support estimate $\hat{\Omega}$ and a probability estimate $\hat{P} = (\hat{p}_1, \ldots, \hat{p}_N)$. The recovery requirements consist of two parts. First, the support set must be exactly recovered with a fixed failure probability $\delta$:

$$\Pr_p\left(\Omega_{obs}\ne\Omega\right)\le\delta,\qquad 0<\delta<\frac{1}{2}.\tag{62}$$

Second, every nonzero probability must satisfy the coordinate-wise relative root-mean-square error requirement

$$\max_{n\in\Omega}\frac{\sqrt{\mathbb{E}_p\left[\left(p_n-p_n\right)^2\right]}}{p_n}\le\varepsilon\,.\tag{63}$$

Here, the expectation is taken over all possible measurement outcomes, including cases in which a particular nonzero index is not observed within the finite sample set; it is not conditioned on successful support recovery.

#### 1) *Quantum Fisher Information and Local Optimality of Index Measurement*

In this subsection, we consider only the local optimality of index projective measurement. This is because quantum Fisher information characterizes the distinguishability of quantum states under infinitesimal perturbations around the true parameter value [32], [33], [34], [35], [36], [37], whereas the quantum Cramér–Rao bound generally applies to locally unbiased estimators or estimators satisfying the corresponding regularity conditions. Therefore, comparing the quantum Fisher information with the classical Fisher information obtained from index measurement can only determine whether the measurement loses information about the probability parameters locally around a given parameter point. Such a comparison alone does not establish global optimality over a finite parameter range, nor does it exclude biased or nonregular estimation methods. We focus here on local optimality to determine whether direct index measurement reaches the local information limit allowed by the quantum state. A uniform lower bound applicable to arbitrary estimators based on index samples will be established in the next subsection using a two-point testing argument.

We first assume that the support set is known and label the nonzero locations as $1, \ldots, K$. The quantum state after domain alignment can be written as

$$|\psi(\mathrm{p})\rangle=\sum_{n=1}^{K}\sqrt{p_n}\,|\omega_n\rangle,\qquad \sum_{n=1}^{K}p_n=1.\tag{64}$$

Because of the normalization constraint on the probabilities, only $K$-1 parameters are independent. Let $\theta = (p_1, \ldots, p_{K-1})^{\mathrm{T}}$, and define

$$p_K=1-\sum_{a=1}^{K-1}p_a\,.\tag{65}$$

For $a = 1, \ldots, K-1$,

$$\frac{\partial|\psi\rangle}{\partial p_a}=\frac{|\omega_a\rangle}{2\sqrt{p_a}}-\frac{|\omega_K\rangle}{2\sqrt{p_K}}\,.\tag{66}$$

At the same time,

$$\left\langle\psi\middle|\frac{\partial\psi}{\partial p_a}\right\rangle=\frac{1}{2}-\frac{1}{2}=0\,.\tag{67}$$

For a pure quantum state, the quantum Fisher information matrix is [32], [33], [34], [35], [36]

$$F_{ab}^{Q}=4\operatorname{Re}\left[\left\langle\frac{\partial\psi}{\partial p_a}\middle|\frac{\partial\psi}{\partial p_b}\right\rangle-\left\langle\frac{\partial\psi}{\partial p_a}\middle|\psi\right\rangle\left\langle\psi\middle|\frac{\partial\psi}{\partial p_b}\right\rangle\right]. \tag{68}$$

Since

$$\left\langle\psi\middle|\frac{\partial\psi}{\partial p_a}\right\rangle=0\,. \tag{69}$$

we obtain

$$F_{ab}^{Q}=\frac{\delta_{ab}}{p_a}+\frac{1}{p_K},\qquad a,b=1,\dots,K-1\,. \tag{70}$$

On the other hand, direct index projective measurement produces an outcome $J$ with probability $\Pr(J=n)=p_n$. The corresponding classical Fisher information matrix is

$$F_{ab}^{J}=\sum_{n=1}^{K}\frac{1}{p_n}\frac{\partial p_n}{\partial p_a}\frac{\partial p_n}{\partial p_b}\,. \tag{71}$$

Using

$$\frac{\partial p_n}{\partial p_a}=-\begin{cases}1, & n=a\\ -1, & n=K\\ 0, & Others\end{cases}\,. \tag{72}$$

we obtain

$$F_{ab}^{J}=\frac{\delta_{ab}}{p_a}+\frac{1}{p_K}\,. \tag{73}$$

Therefore,

$$F^{J}=F^{Q}\,. \tag{74}$$

This shows that, for a fixed support under the real, nonnegative square-root-amplitude model, direct index measurement preserves all quantum Fisher information about the probability parameters [34], [35], [36], [37]. For the probability-estimation task considered here, the index projective measurement after domain alignment therefore reaches the precision limit allowed by the quantum state in the local-information sense. Let $\mathbf{q}=(p_1,\ \dots,\ p_{K-1})^{T}$. Inverting the matrix gives

$$\left(F^{Q}\right)^{-1}=diag(\mathrm{q})-\mathrm{q}\mathrm{q}^{T}\,. \tag{75}$$

For $M_s$ independently prepared quantum states, quantum Fisher information is additive:

$$F_{M_s}^{Q\ J}=M_sF^{Q}\,. \tag{76}$$

Therefore, for any locally unbiased estimator, the quantum Cramér–Rao bound gives [32], [33], [34], [35], [36], [37]

$$Cov\left(\theta\right)\succeq\frac{1}{M_s}\left(F^{Q}\right)^{-1}. \tag{77}$$

From the diagonal elements of the inverse matrix,

$$Var\left(p_n\right)\geq\frac{p_n\left(1-p_n\right)}{M_s}\,. \tag{78}$$

The same result is obtained for $p_K$ by using a symmetric parameterization. Hence, for any nonzero coordinate,

$$\frac{Var\left(p_n\right)}{p_n}\geq\sqrt{\frac{\left(1-p_n\right)}{M_sp_n}}\,. \tag{79}$$

If the relative root-mean-square error of every nonzero probability is required to be no greater than $\varepsilon$, then, within the class of locally unbiased estimators, it is necessary that

$$M_s \geq \frac{1-p_{\min}}{\varepsilon^2 p_{min}}. \tag{80}$$

where $p_{\min} = \min p_n$, $n \in \Omega$. Since every probability distribution supported on $K$ points satisfies $p_{\min} \leq 1/K$, it follows that

$$M_s \geq \frac{K-1}{\varepsilon^2}. \tag{81}$$

Thus, the quantum Cramér–Rao bound yields

$$M_s = \Omega\left(\frac{K}{\varepsilon^2}\right). \tag{82}$$

This result identifies the quantum-statistical origin of the $K/\varepsilon^2$ scaling and also shows that direct index measurement attains the corresponding quantum Fisher information limit. However, the quantum Cramér–Rao bound generally requires local unbiasedness and therefore cannot by itself rule out biased estimators or other nonregular estimation methods. In the next subsection, we establish a uniform lower bound applicable to arbitrary estimators using a two-point testing argument.

2) *Two-Point Testing Lower Bound for Arbitrary Estimators*

The preceding subsection derived the local lower bound $M_s = \Omega(K/\varepsilon^2)$ using the quantum Cramér–Rao bound. However, that result generally requires the estimator to be locally unbiased and to satisfy the corresponding regularity conditions. In this subsection, we continue to assume that the support set is known and fixed as $\Omega = \{1, \ldots, K\}$, and establish a uniform lower bound for arbitrary estimators based on independent index samples using a two-point testing argument.

We use the same two-point family of distributions defined in (25) and (26) of Section IV-A. In Section IV-A, $\alpha$ was taken to be a fixed constant to obtain the basic statistical threshold $M_s = \Omega(K)$ at fixed relative accuracy. Here, instead, $\alpha$ is chosen according to the target accuracy $\varepsilon$ in order to characterize the dependence of the sample complexity on the estimation accuracy. Specifically,

$$p_1^+ = \frac{1+\alpha}{K}, \qquad p_1^- = \frac{1-\alpha}{K}, \tag{83}$$

and

$$p_j^+ = \frac{1-p_1^+}{K-1}, \qquad p_j^- = \frac{1-p_1^-}{K-1}, \qquad j = 2,\ldots,K. \tag{84}$$

When

$$0 < \alpha \leq \min\left\{\frac{1}{4}, 1-c, C-1\right\}, \tag{85}$$

both $p^+$ and $p^-$ belong to the balanced probability class $\mathcal{P}_K(c, C)$. The difference between the two distributions in the first coordinate is

$$p_1^+ - p_1^- = \frac{2\alpha}{K}. \tag{86}$$

Suppose that an estimator $\hat{\mathbf{p}}$ satisfies the coordinate-wise relative root-mean-square error requirement under both distributions. In particular, for the first coordinate,

$$\frac{\sqrt{\mathbb{E}_{\pm}\left[\left(p_1 - p_1^{\pm}\right)^2\right]}}{p_1^{\pm}} \leq \varepsilon, \tag{87}$$

where $\mathbb{E}_+$ and $\mathbb{E}_-$ denote expectations with respect to the index-sample distributions induced by $p^+$ and $p^-$, respectively. We construct the following test according to whether $\hat{p}_1$ lies above or below $1/K$:

$$p_1 > \frac{1}{K} \Rightarrow p^+ , \tag{88}$$

$$p_1 \le \frac{1}{K} \Rightarrow p^- . \tag{89}$$

Under $p^+$, a testing error implies

$$\left| p_1 - p_1^+ \right| \ge \frac{t}{K} . \tag{90}$$

Applying Markov's inequality [31] to the nonnegative random variable $(\hat{p}_1 - p_1^+)^2$ gives

$$\begin{aligned} \Pr\left( p_1 \le \frac{1}{K} \right) &\le \frac{\mathbb{E}_+\left[ \left( p_1 - p_1^+ \right)^2 \right]}{\left( \alpha / K \right)^2} \\ &\le \frac{\varepsilon^2 \left( p_1^+ \right)^2}{\left( \alpha / K \right)^2} \\ &\le \frac{\varepsilon^2 \left( 1+\alpha \right)^2}{\alpha^2} \end{aligned} . \tag{91}$$

Similarly, under $p^-$, a testing error implies

$$\underline{\Pr}\left( p_1 > \frac{1}{K} \right) \le \frac{\varepsilon^2 (1-\alpha)^2}{\alpha^2} . \tag{92}$$

Let $\alpha = 8\varepsilon$ and restrict $0 < \varepsilon \le \varepsilon_0$, where $\varepsilon_0$ is a positive constant depending only on $c$ and $C$, satisfying

$$\varepsilon_0 \le \frac{1}{8} \min\left\{ \frac{1}{4}, 1-c, C-1 \right\} . \tag{93}$$

Under this choice, both testing error probabilities are bounded by fixed constants. Therefore, if an estimator satisfies the coordinate-wise relative root-mean-square error requirement, it must be capable of distinguishing $p^+$ from $p^-$ with a fixed success probability strictly greater than 1/2. The choice $\alpha = 8\varepsilon$ ensures that the separation between the two distributions is of the same order as the estimation-error scale, while guaranteeing a fixed success advantage for the test constructed from the estimator. The constant 8 is not essential; choosing $\alpha = C_0\varepsilon$ for any sufficiently large fixed constant $C_0$ yields the same asymptotic lower bound.

Equations (29)–(35) in Section IV-A already establish the distinguishability lower bound for this two-point family. Specifically, distinguishing $p^+$ from $p^-$ with a fixed success probability requires

$$\frac{M_s \alpha^2}{K} = \Omega(1) . \tag{94}$$

Substituting $\alpha = 8\varepsilon$ gives

$$M_s = \Omega(\frac{K}{\varepsilon^2}) . \tag{95}$$

This lower bound does not require the estimator to be unbiased. The distinction from Section IV-A is that there $\alpha$ was chosen as a fixed constant, yielding only the basic statistical threshold $M_s = \Omega(K)$ at fixed accuracy. Here, $\alpha$ is chosen to be of the same order as the target accuracy $\varepsilon$, and we explicitly establish the implication that an estimator satisfying the relative root-mean-square error requirement can be converted into a reliable two-point test. This yields the explicit $1/\varepsilon^2$ dependence. When $\varepsilon$ is a fixed constant, (95) reduces to $M_s = \Omega(K)$, consistent with the result of Section IV-A.

Sections IV-C1 and IV-C2 also serve different purposes. The former shows that direct index measurement reaches the precision limit permitted by the quantum state in the local Fisher-information sense. The present subsection shows that, within the specified independent index-measurement model, no estimator—including biased and nonregular estimators—can surpass the statistical lower bound in (95).

3) *Joint Lower Bound for Support Recovery and Probability Estimation*

Section IV-B has shown that, under an unknown support, balanced nonzero probabilities, and a fixed failure probability, exact support recovery requires $M_s = \Omega(K \ln K)$. The coordinate-wise probability-estimation lower bound derived in this section is $M_s = \Omega(K/\varepsilon^2)$. Since the same measurement data must simultaneously satisfy both the support-recovery and probability-estimation requirements,

$$M_s = \Omega\left( \max\left\{ K \ln K, \frac{K}{\varepsilon^2} \right\} \right). \tag{96}$$

For any nonnegative $A$ and $B$, $\max\{A, B\} \le A + B \le 2\max\{A, B\}$, and therefore,

$$\max\left\{ K \ln K, \frac{K}{\varepsilon^2} \right\} = \Theta\left( K \ln K + \frac{K}{\varepsilon^2} \right). \tag{97}$$

Hence, the lower bound for the joint recovery task can equivalently be written as

$$M_s = \Omega\left( K \ln K + \frac{K}{\varepsilon^2} \right). \tag{98}$$

Writing the bound as a sum of two terms does not imply that support recovery and probability estimation require two separate sets of samples. The two tasks can share the same set of index samples; the additive form is used only because the maximum and the sum of the two terms are asymptotically equivalent.

4) *Matching Upper Bound and Final Measurement Complexity*

Consider the empirical-frequency estimator $\hat{p}_n = k_n/M_s$, where $k_n$ is the number of occurrences of index $n$ among the $M_s$ measurements. For each $n$ [22], [31],

$$\mathbb{E}_{\mathrm{p}}\left[ p_n \right] = p_n . \tag{99}$$

and

$$Var_{\mathrm{p}}\left( p_n \right) = \frac{p_n (1 - p_n)}{M_s} . \tag{100}$$

Therefore,

$$\frac{\sqrt{\mathbb{E}_{\mathrm{p}}\left[ \left( p_n - p_n \right)^2 \right]}}{p_n} = \sqrt{\frac{1 - p_n}{M_s p_n}} . \tag{101}$$

Let $p_{\min} = \min p_n$, $n \in \Omega$. Then

$$\max_{n\in\Omega} \frac{\sqrt{\mathbb{E}_{\mathrm{p}}\left[ \left( p_n - p_n \right)^2 \right]}}{p_n} \le \sqrt{\frac{1}{M_s p_{\min}}} . \tag{102}$$

Using the balanced-probability condition $p_{\min} \ge c/K$, we obtain

$$\max_{n\in\Omega} \frac{\sqrt{\mathbb{E}_{\mathrm{p}}\left[ \left( p_n - p_n \right)^2 \right]}}{p_n} \le \sqrt{\frac{K}{cM_s}} . \tag{103}$$

Thus, to ensure that the coordinate-wise relative root-mean-square error does not exceed $\varepsilon$, it is sufficient to have

$$M_s \ge \frac{K}{c\varepsilon^2} . \tag{104}$$

For support recovery, we use the observed-support estimator $\hat{\Omega} = \{ n : k_n > 0 \}$. By the union bound,

$$\begin{aligned} \Pr\left( \Omega_{obs} \ne \Omega \right) &\le \sum_{n\in\Omega} (1 - p_n)^{M_s} \\ &\le K \exp(-M_s p_{\min}) . \\ &\le K \exp\left( -\frac{cM_s}{K} \right) \end{aligned} \tag{105}$$

To ensure that the failure probability does not exceed $\delta$, it is sufficient that

$$M_s \geq \frac{K}{c}\ln\left(\frac{K}{\delta}\right). \tag{106}$$

Therefore, a sufficient condition for the same set of samples to satisfy both the support-recovery and probability-estimation requirements is

$$M_s \geq \max\left\{\frac{K}{c}\ln\left(\frac{K}{\delta}\right), \frac{K}{c\varepsilon^2}\right\}. \tag{107}$$

For fixed $\delta$,

$$M_s = O\left(K\ln K + \frac{K}{\varepsilon^2}\right). \tag{108}$$

Combining this upper bound with the joint lower bound yields the Level-III QCS measurement complexity

$$M_s^* = \Theta\left(K\ln K + \frac{K}{\varepsilon^2}\right). \tag{109}$$

Here, the $K \ln K$ term arises from complete coverage of the unknown support set, whereas the $K/\varepsilon^2$ term arises from coordinate-wise relative-accuracy estimation of nonzero probabilities of order $1/K$. When $\varepsilon$ is a fixed constant, $K/\varepsilon^2 = \Theta(K)$, so the support-coverage term dominates, giving

$$M_s^* = \Theta\left(K\ln K\right). \tag{110}$$

When $1/\varepsilon^2 \gg \ln K$ the accuracy term dominates, and

$$M_s^* = \Theta\left(\frac{K}{\varepsilon^2}\right). \tag{111}$$

The two terms are of the same order when

$$\varepsilon \asymp \frac{1}{\sqrt{\ln K}}. \tag{112}$$

The results of this section apply for $2 \leq K \leq N/2$, $0 < \delta < 1/2$, $0 < \varepsilon \leq \varepsilon_0$, where $\delta$ is a fixed constant, the nonzero probabilities satisfy $p_n = \Theta(1/K)$, and the observations are independent single-index projective measurements. If $p_{\min} \ll 1/K$, the numbers of samples required for both support coverage and probability estimation may increase further. In addition, the quantity estimated in this section is the normalized amplitude probability $p_n = |s_n|/||S||_1$. Only when the overall amplitude $||S||_1$ is known does the relation $|s_n| = ||S||_1 p_n$ imply that the relative error in $p_n$ is equal to the relative error in the magnitude of the sparse coefficient. Recovery of the coefficient sign, phase, and unknown overall scale is not included in the recovery task considered in this section.

## V. Conversion Between the Number of Effective Index Samples and the Number of Experimental Rounds

The measurement complexities discussed in Section IV are all expressed in terms of the total number of effective index samples $M_s$. In this section, we further distinguish the total number of effective index samples $M_s$ from the number of experimental rounds $M$, and analyze their relationship when each experimental round produces $L$ effective indices. It is important to emphasize that reducing the number of experimental rounds does not imply a reduction in the total amount of information required for statistical recovery. The lower bounds derived in Section IV constrain $M_s$, whereas multi-index readout changes the information throughput per experimental round.

### A. *Effective Index Samples and Experimental Rounds*

Under standard single-index projective measurement, each experimental round consists of one quantum-state preparation, one domain-alignment evolution, and one projective measurement, producing a single index outcome. Therefore, $L = 1$ and $M_s = M$. If one experimental round can produce $L$ effective index outcomes, the outputs of the $r$-th round can be written as $J_{r,1}, J_{r,2}, \ldots, J_{r,L}$, $r = 1, \ldots, M$. Here, an "effective index" means that each outcome follows the probability distribution specified in Section IV: $\Pr(J_{r,l} = n) = p_n$, $l = 1, \ldots, L$. If the $L$ indices within the same round are mutually independent conditioned on $\mathbf{p}$, and different experimental rounds are also mutually independent, then the total number of effective index samples is $M_s = ML$. Thus, $M_s$ represents the total number of independent index samples actually used for statistical recovery, whereas $M$ represents the number of times the experiment is repeated. The two quantities are equal only when $L$=1.

### B. Multi-Index Readout Model with Fixed L

With multimode photon-number-resolving detection, a single experimental round can output the photon counts in the individual measurement modes [38], [39], [40], [41]. Let $C_r = (C_{r,1}, \ldots, C_{r,N})$ denote the count vector in the $r$-th round, where $C_{r,n}$ is the number of effective events detected in the mode corresponding to index $n$. Suppose that each round produces exactly $L$ effective indices. Then

$$\sum_{n=1}^{N} C_{r,n} = L . \tag{113}$$

If these events are equivalent to $L$ independent samples drawn from the probability distribution $\mathbf{p}$, then $C_r \sim$ Multinomial $(L; p_1, \ldots, p_N)$. More explicitly, for any count vector satisfying

$$c_{r,n} \geq 0, \quad \sum_{n=1}^{N} c_{r,n} = L . \tag{114}$$

We have [22], [38]

$$\Pr\left(C_r = c_r\right) = \frac{L!}{\prod_{n=1}^{N} c_{r,n}!} \prod_{n=1}^{N} p_n^{c_{r,n}} . \tag{115}$$

After $M$ experimental rounds, the total count of index $n$ is

$$k_n = \sum_{r=1}^{M} C_{r,n} . \tag{116}$$

Because different experimental rounds are mutually independent and each round follows the same probability distribution $\mathbf{p}$, the total count vector satisfies $(k_1, \ldots, k_N) \sim$ Multinomial $(ML; p_1, \ldots, p_N)$. Defining $M_s = ML$, this can equivalently be written as $(k_1, \ldots, k_N) \sim$ Multinomial $(M_s; p_1, \ldots, p_N)$. Therefore, under the above independence assumptions, $M$ experimental rounds with $L$ indices per round have exactly the same statistical model as $M_s = ML$ standard single-index samples. The corresponding empirical-probability estimator is

$$p_n = \frac{k_n}{ML} = \frac{k_n}{M_s} . \tag{117}$$

Its expectation and variance are, respectively,

$$\mathbb{E}\left[p_n\right] = p_n . \tag{118}$$

and

$$Var\left(p_n\right) = \frac{p_n\left(1 - p_n\right)}{ML} = \frac{p_n\left(1 - p_n\right)}{M_s} . \tag{119}$$

Thus, the probability-estimation accuracy is determined solely by the total number of effective index samples $M_s$, rather than by how these samples are distributed among experimental rounds. Similarly, the classical Fisher information provided by one experimental round with multi-index readout is

$$F^{(L)} = LF^{J} , \tag{120}$$

where $F^J$ is the classical Fisher information provided by a single index sample [35], [36]. After $M$ experimental rounds, $F^{(M,L)} = MLF^J = M_sF^J$. Therefore, increasing $L$ increases the Fisher information obtained per experimental round, but does not change the amount of information provided by each effective index sample.

### C. Experimental Rounds for the Three Recovery Levels

#### 1) Basic Statistical Information

Section IV-A established the first-level necessary condition $M_s = \Omega(K)$. Substituting $M_s = ML$ gives $ML = \Omega(K)$, and therefore the number of experimental rounds must satisfy $M = \Omega(K/L)$. Since at least one experimental round is required, a more precise expression is

$$M = \Omega\left(\max\left\{1, \frac{K}{L}\right\}\right). \tag{121}$$

This result is only a necessary condition for basic statistical estimation at fixed accuracy and is not the tight complexity for exact recovery of an unknown support.

2) *Exact Recovery of an Unknown Support*

Section IV-B showed that, under the balanced probability class and a fixed support-recovery failure probability, exact recovery of an unknown support requires $M_s^* = \Theta(K \ln K)$. If each experimental round provides $L$ independent effective indices, then

$$M^* = \left\lceil \frac{M_s^*}{L} \right\rceil. \tag{122}$$

Therefore,

$$M^* = \Theta\left(\max\left\{1, \frac{K \ln K}{L}\right\}\right). \tag{123}$$

This conversion does not eliminate the $K \ln K$ effective index samples required for coverage; rather, it distributes these samples across experimental rounds, with $L$ effective indices produced per round. To provide an explicit achievable upper bound for the observed-support estimator, let $p_{\min} = \min p_n$, $n \in \Omega$. After $M$ experimental rounds, each nonzero index has $ML$ opportunities to be sampled. Therefore, the support-recovery failure probability satisfies

$$\Pr\left(\Omega_{obs} \neq \Omega\right) \leq K \exp(-MLp_{\min}) . \tag{124}$$

For the balanced probability class, $p_{\min} \geq c/K$, and hence

$$\Pr\left(\Omega_{obs} \neq \Omega\right) \leq K \exp\left(-\frac{cML}{K}\right). \tag{125}$$

To ensure that the failure probability does not exceed $\delta$, it is sufficient that

$$ML \geq \frac{K}{c} \ln\left(\frac{K}{\delta}\right). \tag{126}$$

Equivalently,

$$M \geq \frac{K}{cL} \ln\left(\frac{K}{\delta}\right). \tag{127}$$

Accounting for the integer-valued number of experimental rounds, a sufficient condition is

$$M \geq \left\lceil \frac{K}{cL} \ln\left(\frac{K}{\delta}\right) \right\rceil. \tag{128}$$

3) *Support Recovery with Coordinate-Wise Accuracy*

Section IV-C established that the total effective index sample complexity required for the third-level recovery task is

$$M_s^* = \Theta\left(K \ln K + \frac{K}{\varepsilon^2}\right). \tag{129}$$

If each experimental round produces $L$ mutually independent effective indices, then the minimum number of experimental rounds required to complete this task is

$$M^* = \left\lceil \frac{M_s^*}{L} \right\rceil. \tag{130}$$

Since at least one experimental round is required, the asymptotic scaling can be written as

$$M^* = \Theta\left[\max\left\{1, \frac{K \ln K + K / \varepsilon^2}{L}\right\}\right]. \tag{131}$$

Here, the $K \ln K/L$ contribution corresponds to the number of experimental rounds required to completely cover the unknown support set, whereas $K/(L\varepsilon^2)$ corresponds to the number of rounds required for coordinate-wise relative-accuracy estimation. Increasing $L$ can reduce the number of experimental rounds proportionally, but it does not change the total number of effective index samples required by the recovery task.

*D. Effect of L on the Number of Experimental Rounds*

Let the total number of effective index samples required for the Level-III recovery task be

$$A(K,\varepsilon) = K \ln K + \frac{K}{\varepsilon^2}. \tag{132}$$

Then the number of experimental rounds can be written uniformly as

$$M^* = \Theta\left[\max\left\{1, \frac{A(K,\varepsilon)}{L}\right\}\right]. \tag{133}$$

This result shows that, under the independent effective-index model, $M$ decreases approximately as $1/L$ as $L$ increases, but can decrease only to a constant order. When each experimental round produces only one index, $L = 1$, and hence $M = M_s$. The Level-III complexity is therefore

$$M^* = \Theta\left[K \ln K + \frac{K}{\varepsilon^2}\right]. \tag{134}$$

This corresponds to the standard single-index projective-measurement model considered in Section IV.

When $1 \ll L \ll A(K, \varepsilon)$, each experimental round produces multiple effective indices, but $L$ remains smaller than the total sample requirement. In this regime,

$$M^* = \Theta\left[\frac{A(K,\varepsilon)}{L}\right]. \tag{135}$$

Thus, doubling $L$ approximately halves the number of experimental rounds in the asymptotic sense. Accordingly, $L$ characterizes the information throughput per experimental round.

When $L = \Theta(K)$, the number of effective indices produced per round is of the same order as the sparsity level. If $\varepsilon$ is a fixed constant, then $A(K, \varepsilon) = \Theta[K \ln K]$, and hence

$$M^* = \Theta(\ln K). \tag{136}$$

Therefore, producing $\Theta(K)$ effective indices per round reduces the number of experimental rounds from $\Theta(K \ln K)$ to $\Theta(\ln K)$, while the total number of effective index samples remains $\Theta(K \ln K)$. If the accuracy term dominates, i.e.,

$$\frac{1}{\varepsilon^2} \gg \ln K, \tag{137}$$

then

$$A(K,\varepsilon) = \Theta\left[\frac{K}{\varepsilon^2}\right]. \tag{138}$$

For $L = \Theta(K)$,

$$M^* = \Theta\left[\frac{1}{\varepsilon^2}\right]. \tag{139}$$

When $\varepsilon$ is a fixed constant and $L = \Theta(K \ln K)$, we have $M^* = \Theta(1)$. This means that one or a constant number of experimental rounds can provide all the effective index samples required for support coverage. However, this does not imply that the total measurement resource becomes constant, because each experimental round already contains $\Theta(K \ln K)$ effective detection events.

When $L \geq A(K, \varepsilon)$, that is,

$$L \geq K \ln K + \frac{K}{\varepsilon^2}, \tag{140}$$

we have $M^* = \Theta(1)$. The number of experimental rounds cannot continue to decrease as $1/L$, because at least one experimental round must be performed. Thus, the reduction of $M$ with increasing $L$ eventually saturates: $M^* \geq 1$.

### *E. Applicability of the Conversion Relation*

The key condition for the relation $M_s = ML$ to hold is that the $L$ outcomes produced in each experimental round are statistically equivalent to $L$ independent effective index samples. Photon-number-resolving detection alone does not automatically guarantee this condition. If the multiphoton outputs within the same experimental round exhibit non-negligible quantum correlations, bunching effects, or other correlations, the count vector does not necessarily follow Multinomial($L$; $p$). In such cases, each detection event cannot be directly counted as an independent effective index sample [38], [39], [40], [41]. Instead, the effective sample size should be redefined according to the actual joint probability distribution, covariance structure, or Fisher information.

If the number of effective indices produced in the $r$-th experimental round is not fixed but is instead a random variable $L_r$, then the actual total number of effective index samples is

$$M_s = \sum_{r=1}^{M} L_r \, . \tag{141}$$

Conditioned on $L_1, \ldots, L_M$, if all indices remain mutually independent, the total count vector satisfies

$$\left(k_1,\ldots,k_N\right) \middle| \left(L_1,\ldots,L_M\right) \sim Multinomial\left(\sum_{r=1}^{M} L_r ; p_1,\ldots,p_N\right). \tag{142}$$

In this case, whether the required sample budget for recovery has been reached should be determined from the actually accumulated $M_s$, rather than from the average number of photons per round alone. If only $\mathbb{E}[L_r] = \bar{L}$ is known, then $\mathbb{E}[M_s] = M\bar{L}$, but this is only an expectation-level relation. To guarantee a fixed success probability, the lower-tail fluctuations of $M_s$ must also be controlled.

Detection loss must likewise be reflected in the effective value of $L$. If $L_{in}$ probes are generated in one round but only $L_{det}$ yield usable index outcomes, then $L = L_{det}$ should be used rather than the number of input probes $L_{in}$. If the detection efficiencies differ across modes, the actual observed probability becomes

$$p_n = \frac{\eta_n p_n}{\sum_{j=1}^{N} \eta_j p_j} , \tag{143}$$

where $\eta_n$ is the detection efficiency of the mode corresponding to index $n$. Unless the $\eta_n$ are known and properly calibrated, mode-dependent loss introduces systematic bias into probability estimation. In summary, multi-index readout or photon-number-resolving detection can reduce the number of experimental rounds to

$$M^* = \Theta\left[\max\left\{1, \frac{K \ln K + K/\varepsilon^2}{L}\right\}\right], \tag{144}$$

but does not change the total effective index sample complexity,

$$M_s^* = \Theta\left(K \ln K + \frac{K}{\varepsilon^2}\right). \tag{145}$$

Therefore, increasing $L$ improves the information throughput per experimental round rather than overcoming the statistical sample lower bounds derived in Section IV. Comparing only the number of experimental rounds without simultaneously reporting the number of effective indices per round may incorrectly interpret the batching benefit of parallel readout as a reduction in the total measurement resource.

## VI. Conclusion

This work investigates the measurement resources required by QCS for sparse-support identification and amplitude estimation, as well as the origin of its difference from classical non-adaptive linear CS, under the common conditions of a known sparse basis, an unknown support set, non-adaptive measurements, no additional prior information, finite measurement resources, and a common recovery criterion.

For classical non-adaptive linear compressed sensing, we take exact recovery of an unknown support set as the core comparison task. Under a finite per-measurement information budget or dimension-independent stable-embedding conditions, the number of measurements satisfies the information-theoretic lower bound

$$M = \Omega\left(K \log \frac{N}{K}\right). \tag{146}$$

We then characterize the resources required by QCS in terms of both the total number of effective index samples $M_s$

and the number of experimental rounds $M$, assuming that domain alignment and index-resolved readout can be physically implemented. Under standard independent index measurements, three levels of results are obtained.

First, obtaining basic statistical information about the $K$ nonzero components requires at least $M_s = \Omega(K)$. This is only a coarse necessary condition at fixed accuracy and does not require exact recovery of the unknown support.

Second, when the support is unknown, the nonzero probabilities satisfy $p_n = \Theta(1/K)$, and exact support recovery is required with a fixed success probability,

$$M_s^* = \Theta\left(K \ln K\right). \tag{147}$$

This result uses the same support-recovery objective and success-probability requirement as those adopted for classical CS and therefore constitutes the core result for the classical–quantum comparison in this work.

Third, if, in addition to exact support recovery, the coordinate-wise relative root-mean-square error of each nonzero probability is required to be no greater than $\varepsilon$, then

$$M_s^* = \Theta\left(K \ln K + \frac{K}{\varepsilon^2}\right), \tag{148}$$

where the $K \ln K$ term arises from the complete-coverage requirement under independent sampling with replacement, whereas the $K/\varepsilon^2$ term arises from coordinate-wise relative-accuracy estimation of nonzero probabilities of order $1/K$.

It should be emphasized that the third level is an extension of the core support-recovery task with an additional amplitude-accuracy requirement and should not be directly compared with the classical CS lower bound in Section II, which concerns support recovery alone. If the same coordinate-wise amplitude-error requirement were imposed on classical CS, its measurement lower bound would likewise incur an additional cost depending on the accuracy parameter $\varepsilon$. The precise form of this accuracy-dependent term depends on the noise, per-measurement information budget, and amplitude-observation model of the classical measurement scheme and is not derived here. Accordingly, the direct comparison between QCS and classical CS in this work is based on the second-level support-recovery task.

If each experimental round produces $L$ mutually independent effective indices, then $M_s = ML$, and the corresponding number of experimental rounds for the third-level task is

$$M = \Theta\left(\max\left\{1, \frac{K \ln K + K/\varepsilon^2}{L}\right\}\right). \tag{149}$$

Thus, multi-index readout can increase the information throughput per experimental round and reduce the number of experimental rounds, but it does not change the total effective index sample complexity required for recovery.

These results show that the conditional advantage of QCS arises from the change in how positional information is encoded after domain alignment: the support locations are directly carried by the index labels of the measurement outcomes, rather than being inferred from classical continuous-valued measurements. This result does not imply that QCS unconditionally outperforms all classical schemes, nor does it constitute a universal information-theoretic lower bound for arbitrary quantum measurements. Its scope is restricted to the signal-access and index-readout models specified in this work and relies on a known sparse basis, implementable domain alignment, an unknown and unstructured support set, balanced nonzero probabilities, independent index sampling, and a fixed failure probability. The recovery task considered in this work includes only the magnitudes of the normalized sparse coefficients and does not address their signs, phases, or unknown overall scale. Correlated outputs, loss, dark counts, device complexity, and possible improvements enabled by entangled or coherent joint measurements are also outside the scope of the present results.

## REFERENCES

[1] C. E. Shannon, “A mathematical theory of communication,” Bell Syst. Tech. J., vol. 27, pp. 379–423, 623–656, Jul. 1948, doi: 10.1002/j.1538-7305.1948.tb01338.x.

[2] H. Nyquist, “Certain topics in telegraph transmission theory,” *Trans. Amer. Inst. Electr. Eng.*, vol. 47, no. 2, pp. 617–644, Apr. 1928, doi: 10.1109/T-AIEE.1928.5055024.

[3] A. J. Jerri, “The Shannon sampling theorem—its various extensions and applications: a tutorial review,” *Proc. IEEE*, vol. 65, no. 11, pp. 1565–1596, Nov. 1977, doi: 10.1109/PROC.1977.10771.

[4] D. L. Donoho, “Compressed sensing,” *IEEE Trans. Inf. Theory*, vol. 52, no. 4, pp. 1289–1306, Apr. 2006, doi: 10.1109/TIT.2006.871582.

[5] E. J. Candès, J. Romberg, and T. Tao, “Robust uncertainty principles: exact signal reconstruction from highly incomplete frequency information,” *IEEE Trans. Inf. Theory*, vol. 52, no. 2, pp. 489–509, Feb. 2006, doi: 10.1109/TIT.2005.862083.

[6] E. J. Candès and T. Tao, “Decoding by linear programming,” *IEEE Trans. Inf. Theory*, vol. 51, no. 12, pp. 4203–4215, Dec. 2005, doi: 10.1109/TIT.2005.858979.

[7] E. J. Candès and T. Tao, “Near-optimal signal recovery from random projections: Universal encoding strategies?,” *IEEE Trans. Inf. Theory*, vol. 52, no. 12, pp. 5406–5425, Dec. 2006, doi: 10.1109/TIT.2006.885507.

[8] E. J. Candès, “The restricted isometry property and its implications for compressed sensing,” *C. R. Math.*, vol. 346, no. 9–10, pp. 589–592, 2008, doi: 10.1016/j.crma.2008.03.014.
[9] R. Baraniuk, M. Davenport, R. DeVore, and M. Wakin, “A simple proof of the restricted isometry property for random matrices,” *Constr. Approx.*, vol. 28, no. 3, pp. 253–263, Dec. 2008, doi: 10.1007/s00365-007-9003-x.
[10] S. Foucart and H. Rauhut, *A mathematical introduction to compressive sensing*. New York, NY, USA: Birkhäuser/Springer, 2013. doi: 10.1007/978-0-8176-4948-7.
[11] A. Cohen, W. Dahmen, and R. DeVore, “Compressed sensing and best $k$-term approximation,” *J. Amer. Math. Soc.*, vol. 22, no. 1, pp. 211–231, 2009, doi: 10.1090/S0894-0347-08-00610-3.
[12] M. J. Wainwright, “Information-theoretic limits on sparsity recovery in the high-dimensional and noisy setting,” *IEEE Trans. Inf. Theory*, vol. 55, no. 12, pp. 5728–5741, Dec. 2009, doi: 10.1109/TIT.2009.2032816.
[13] A. K. Fletcher, S. Rangan, and V. K. Goyal, “Necessary and sufficient conditions for sparsity pattern recovery,” *IEEE Trans. Inf. Theory*, vol. 55, no. 12, pp. 5758–5772, Dec. 2009, doi: 10.1109/TIT.2009.2032726.
[14] W. Wang, M. J. Wainwright, and K. Ramchandran, “Information-theoretic limits on sparse signal recovery: Dense versus sparse measurement matrices,” *IEEE Trans. Inf. Theory*, vol. 56, no. 6, pp. 2967–2979, Jun. 2010, doi: 10.1109/TIT.2010.2046199.
[15] M. Akçakaya and V. Tarokh, “Shannon-theoretic limits on noisy compressive sampling,” *IEEE Trans. Inf. Theory*, vol. 56, no. 1, pp. 492–504, Jan. 2010, doi: 10.1109/TIT.2009.2034796.
[16] S. Aeron, V. Saligrama, and M. Zhao, “Information theoretic bounds for compressed sensing,” *IEEE Trans. Inf. Theory*, vol. 56, no. 10, pp. 5111–5130, Oct. 2010, doi: 10.1109/TIT.2010.2059891.
[17] G. Reeves and M. C. Gastpar, “Approximate sparsity pattern recovery: Information-theoretic lower bounds,” *IEEE Trans. Inf. Theory*, vol. 59, no. 6, pp. 3451–3465, Jun. 2013, doi: 10.1109/TIT.2013.2253852.
[18] J. Hu *et al.*, “Quantum compressed sensing,” 2026, *arXiv*: 2605.15784.
[19] J. Hu *et al.*, “Quantum compressed sensing of ultra-wideband radio-frequency signal,” 2021, *arXiv*: 2106.13668.
[20] L. Yang *et al.*, “Three-dimensional quantum imaging of dynamic targets using quantum compressed sensing,” *Opt. Express*, vol. 32, no. 4, pp. 6025–6036, Feb. 2024, doi: 10.1364/OE.517160.
[21] S. Wu *et al.*, “10-km passive drone detection using broadband quantum compressed sensing imaging,” *Light Sci. Appl.*, vol. 14, Art. no. 244, Jul. 2025, doi: 10.1038/s41377-025-01878-y.
[22] T. M. Cover and J. A. Thomas, *Elements of information theory*, 2nd ed. Hoboken, NJ, USA: Wiley, 2006.
[23] R. M. Fano, *Transmission of information; a statistical theory of communications*. Cambridge, MA, USA: MIT Press, 1961.
[24] A. B. Tsybakov, *Introduction to nonparametric estimation*. New York, NY, USA: Springer, 2009. doi: 10.1007/b13794.
[25] M. A. Nielsen and I. L. Chuang, *Quantum computation and quantum information*, 10th Anniversary ed. Cambridge, U.K.: Cambridge Univ. Press, 2010.
[26] A. S. Holevo, “Bounds for the quantity of information transmitted by a quantum communication channel,” *Probl. Inf. Transm.*, vol. 9, no. 3, pp. 177–183, 1973.
[27] L. Le Cam, *Asymptotic methods in statistical decision theory*. New York, NY, USA: Springer-Verlag, 1986.
[28] M. S. Pinsker, *Information and information stability of random variables and processes*. San Francisco, CA, USA: Holden-Day, 1964.
[29] P. Erdős and A. Rényi, “On a classical problem of probability theory,” *Magyar Tudományos Akadémia Matematikai Kutató Intézetének Közleményei*, vol. 6, no. 1–2, pp. 215–220, 1961.
[30] A. Boneh and M. Hofri, “The coupon-collector problem revisited — a survey of engineering problems and computational methods,” *Stochastic Models*, vol. 13, no. 1, pp. 39–66, Jan. 1997, doi: 10.1080/15326349708807412.
[31] M. Mitzenmacher and E. Upfal, *Probability and computing: Randomization and probabilistic techniques in algorithms and data analysis*, 2nd ed. Cambridge, U.K.: Cambridge Univ. Press, 2017.
[32] C. W. Helstrom, *Quantum detection and estimation theory*. New York, NY, USA: Academic Press, 1976.
[33] A. S. Holevo, *Probabilistic and statistical aspects of quantum theory*. Amsterdam, The Netherlands: North-Holland, 1982.
[34] S. L. Braunstein and C. M. Caves, “Statistical distance and the geometry of quantum states,” *Phys. Rev. Lett.*, vol. 72, no. 22, pp. 3439–3443, May 1994, doi: 10.1103/PhysRevLett.72.3439.
[35] M. G. A. Paris, “Quantum estimation for quantum technology,” *Int. J. Quantum Inf.*, vol. 7, suppl. 1, pp. 125–137, Jan. 2009, doi: 10.1142/S0219749909004839.
[36] J. Liu, H. Yuan, X.-M. Lu, and X. Wang, “Quantum fisher information matrix and multiparameter estimation,” *J. Phys. A, Math. Theor.*, vol. 53, no. 2, Art. no. 023001, 2020, doi: 10.1088/1751-8121/ab5d4d.
[37] R. D. Gill and S. Massar, “State estimation for large ensembles,” *Phys. Rev. A*, vol. 61, no. 4, Art. no. 042312, Mar. 2000, doi: 10.1103/PhysRevA.61.042312.
[38] L. Mandel and E. Wolf, *Optical coherence and quantum optics*. Cambridge, U.K.: Cambridge Univ. Press, 1995.
[39] R. H. Hadfield, “Single-photon detectors for optical quantum information applications,” *Nat. Photonics*, vol. 3, no. 12, pp. 696–705, Dec. 2009, doi: 10.1038/nphoton.2009.230.
[40] M. J. Fitch, B. C. Jacobs, T. B. Pittman, and J. D. Franson, “Photon-number resolution using time-multiplexed single-photon detectors,” *Phys. Rev. A*, vol. 68, no. 4, Art. no. 043814, Oct. 2003, doi: 10.1103/PhysRevA.68.043814.
[41] D. Achilles *et al.*, “Photon-number-resolving detection using time-multiplexing,” *J. Mod. Opt.*, vol. 51, no. 9–10, pp. 1499–1515, Jun. 2004, doi: 10.1080/09500340408235288.